\documentclass[a4paper, amsfonts, amssymb, amsmath, reprint, prl, showkeys, superscriptaddress, floatfix]{revtex4-1}

\usepackage{amsthm}
\usepackage{mathtools}
\usepackage{physics}
\usepackage{xcolor}
\usepackage{graphicx}
\usepackage{adjustbox}
\usepackage{placeins}
\usepackage[T1]{fontenc}
\usepackage[utf8]{inputenc}
\usepackage{csquotes}
\usepackage{booktabs}

\usepackage{soul} 

\usepackage{journals}

\usepackage[pdftitle={Article}, pdfauthor={Author}]{hyperref} 

\newcommand{\ie}{i.e.,~}
\newcommand{\eg}{e.g.,~}
\newcommand{\unit}[1]{%
    \ensuremath{\, \mathrm{#1}}}

\begin{document}
\title{Fast Rotating Relativistic Stars: Spectra and Stability without Approximation}

\author{Christian J. Kr\"uger}
    \email{christian.krueger@tat.uni-tuebingen.de}
    \affiliation{Theoretical Astrophysics, IAAT, University of T\"ubingen, 72076 T\"ubingen, Germany}
    \affiliation{Department of Physics, University of New Hampshire, 9 Library Way, Durham, NH 03824, USA}
\author{Kostas D. Kokkotas}
    \email{kostas.kokkotas@uni-tuebingen.de}
    \affiliation{Theoretical Astrophysics, IAAT, University of T\"ubingen, 72076 T\"ubingen, Germany}

\date{\today} 

\begin{abstract}
We study oscillations and instabilities of relativistic stars using perturbation theory in general relativity and take into account the contribution of a dynamic spacetime. We present the oscillation spectrum as well as the critical values for the onset of the secular CFS instability of neutron stars, and propose universal relations for gravitational wave asteroseismology, which may help constrain the neutron star radius and/or the nuclear equation of state. The results are relevant for all stages during a neutron star's life but especially to nascent or remnant objects following a binary merger.
\end{abstract}


\maketitle

\emph{Introduction.}---Oscillations and instabilities of neutron stars were always considered among the promising sources for gravitational waves. This was the reason that attracted the interest of Kip Thorne and collaborators \cite{1967ApJ...149..591T,1969ApJ...158....1T} already from the mid-1960s, while instabilities associated with spinning neutron stars, \eg the so-called CFS instability, discovered and analyzed by Chandrasekhar \cite{1970PhRvL..24..611C} with Friedman and Schutz \cite{1975ApJ...199L.157F,1978ApJ...222..281F}, are still promising sources. Rotational instabilities, such as the CFS instability, can be excited in core collapse scenarios \cite{2006ApJS..164..130O} and they can be potential sources of gravitational waves in the late postmerger phase \cite{2015PhRvD..92j4040D}, \ie when the final object survives for periods longer than a few tenths of seconds. Furthermore, the CFS instability of the $r$-mode can be active in low-mass x-ray binaries \cite{2016EPJA...52...38K} and  even affect the evolution of single neutron stars as proposed recently \cite{2017PhRvL.119p1103H, 2018ApJ...864..137A, 2020ApJ...895...11F}. 

The different oscillation patterns of a neutron star are characterized by their restoring force, \eg $p$(pressure)-modes, $g$(gravity)-modes, $i$(Coriolis)-modes, $s$(shear)-modes or $w$(wave)-modes. The $f$-mode is the fundamental mode of the $p$-mode sequence and it is the oscillation mode most likely to be excited in violent processes such as neutron star mergers or neutron star formation by supernova core collapse \cite{2017ApJ...837...67C, 2018PhRvD..98j4005C, 2018MNRAS.474.5272T, 2019MNRAS.482.3967T}. The $f$-mode is associated with major density variations and thus can potentially be an emitter of copious amounts of gravitational radiation. The emission of gravitational waves is the primary reason for the mode's rapid damping at least for newly born neutron stars.

The efforts to associate the patterns of oscillations with the bulk parameters of the stars, \eg their mass, radius or equation of state (henceforth EoS) was initiated in the mid-1990s and continued for almost two decades, advancing the field of gravitational wave asteroseismology \cite{1996PhRvL..77.4134A, 1998MNRAS.299.1059A, 2001MNRAS.320..307K, 2004PhRvD..70l4015B, 2005ApJ...629..979L, 2005PhRvL..95o1101T, 2010ApJ...714.1234L}. To date, very robust empirical relations have been derived for non-rotating neutron stars, connecting observables such as frequency, damping time, or moment of inertia $I$ to the bulk properties; for example, relations of the form $\sigma_0 = \alpha + \beta \sqrt{ M_0/R_0^3 }$ or $M\sigma_0 = F(M_0^3/I)$ (cf. \cite{1996PhRvL..77.4134A, 2010ApJ...714.1234L}) could provide the average density or the moment of inertia of the star if the $f$-mode frequency $\sigma_0$ is known.

In the era of gravitational wave astronomy, the various oscillation patterns (traced already in numerical simulations, \eg \cite{Baiotti08,2011MNRAS.418..427S,2015PhRvL.115i1101B}), if observed, can provide a wealth of information about the emitting sources and their effects can leave their imprints both in the gravitational but also in the electromagnetic spectrum. Moreover, recent studies relate the $f$-mode frequencies to the Love numbers \cite{2019PhRvC..99d5806W, 2020NatCo..11.2553P, 2019arXiv190500012A} and even to the postmerger short gamma-ray bursts \cite{2019ApJ...884L..16C}.

In these early works the rotation of neutron stars was not taken into account within a relativistic framework but its effect was typically extrapolated from Newtonian studies. In nature, neutron stars will always rotate and their rotation rate may reach extreme values. In fact, from the point of view of gravitational wave detectability of oscillation modes, the most relevant scenarios are likely to involve rapidly rotating stars. Unfortunately, the aforementioned empirical relations cannot be trivially extended to rotating stars. Rotation splits the oscillation spectra in a similar fashion as the Zeeman splitting of the spectral lines due to the presence of magnetic fields. In rotating stars, the splitting leads to perturbations propagating in the direction of rotation (so-called \emph{co}-rotating modes) and perturbations traveling in the opposite direction (\emph{counter}-rotating modes). The oscillation frequency as observed by an observer at infinity will either increase or decrease depending on the propagation direction of the waves; for slow rotation there will be a shift of the form $\sigma = \sigma_0 \pm \kappa m \Omega + \mathcal{O}(\Omega^2)$ where $m$ is the angular harmonic index, $\kappa$ a mode-dependent constant and $\Omega$ the angular rotation rate of the star.  If the spin of the star exceeds a critical value, which depends on, \eg the EoS and its mass---\ie when the pattern velocity $\sigma/m$ of the backward moving mode becomes smaller than the star's rotation rate $\Omega/2\pi$---then the star becomes unstable to the emission of gravitational radiation; this is the aforementioned CFS instability \cite{1970PhRvL..24..611C, 1975ApJ...199L.157F, 1978ApJ...222..281F}. This instability is generic (independent of the degree of rotation) for the $r$-modes \cite{1998ApJ...502..708A, 1998ApJ...502..714F} while it can be excited only for relatively high spin values ($\Omega \gtrapprox 0.8 \Omega_K$, with $\Omega_K$ the Kepler velocity) for the quadrupolar $f$-modes. An extensive discussion can be found in \cite{2017LRR....20....7P, 2018ASSL..457..673G}.

For the sake of clarity, we will use upper indices ``s'' and ``u'' on the coefficients of our models to distinguish between the \textbf{s}table (co-rotating) and the potentially \textbf{u}nstable (counter-rotating) branch of the $f$-mode.

Throughout this Letter, we employ units in which $c=G=M_\odot=1$.

\emph{Mathematical formulation.}---The mathematical formulation of the problem and the technical description of the time evolution code we developed, along with convergence tests, are laid out along with all relevant details in an companion paper \cite{2020PhRvD.102f4026K}; nonetheless, we repeat the fundamentals here for completeness.

We work with the Einstein equations along with the law for the conservation of energy-momentum,
\begin{equation}
    G_{\mu\nu} = 8\pi T_{\mu\nu}
    \quad\text{and}\quad
    \nabla_\mu T^{\mu\nu} = 0.
    \label{eq:Einstein}
\end{equation}
We restrict ourselves to the study of small perturbations around equilibrium configurations which allows us to linearize Eq. \eqref{eq:Einstein}. The stationary and axisymmetric background configuration is described in quasi-isotropic coordinates of the form
\begin{align}
    ds^2 & = - e^{2\nu} dt^2 + e^{2\psi} r^2 \sin^2 \theta
             (d\phi - \omega dt)^2 \nonumber \\
         & \qquad + e^{2\mu} (dr^2 + r^2 d\theta^2),
           \label{eq:metric}
\end{align}
and the neutron star is modelled as a perfect fluid for which the corresponding energy-momentum tensor  takes the form
\begin{equation}
    T^{\mu\nu} = (\epsilon + p) u^\mu u^\nu + p g^{\mu\nu},
    \label{eq:Energy-Momentum}
\end{equation}
where $\epsilon$ is the energy density, $p$ is the pressure, and $u^\mu$ is the 4-velocity of the fluid. We generate equilibrium configurations of uniformly rotating neutron stars by means of the \texttt{rns}-code \cite{1995ApJ...444..306S, 1998A&AS..132..431N, rns-v1.1}. An EoS links the energy density and the pressure to each other; while our focus is on realistic EoSs, we will for the purposes of comparison also utilize commonly used polytropic EoSs~\cite{1965ApJ...142.1541T}. As in previous studies, we consider sequences of neutron stars along which we keep the central energy density constant, with rotation rates up to their respective mass-shedding limit; we choose the three different polytropic indices $N = 0.6849, 0.7463$, and $1$. Beside the polytropic EoSs, we employ piecewise-polytropic approximations \cite{2009PhRvD..79l4032R} to the four tabulated EoSs (APR4, H4, SLy, and WFF1) for the scrutiny of more realistic neutron star models, for which we also generate rotational sequences of fixed baryon mass. Our nonrotating configurations have gravitational masses $M \in [1.13,\,2.19] M_\odot$. Even though current astrophysical constraints play a role in our particular choice of EoSs, it is largely motivated by our desire to provide robust universal relations by covering a wide part of the parameter space.

\emph{Results.}---As shown in the companion paper \cite{2020PhRvD.102f4026K}, our code produces results in excellent agreement with previously published values \cite{2010PhRvD..81h4055Z, 2018PhRvD..98j4005C} and our convergence tests demonstrate an accuracy of the obtained frequencies of $1 - 2\%$. In this Letter, we will provide some highlighted results in order to demonstrate the existence of asteroseismological relations of various types and we lay out the way that one can make use of these relations in analyzing gravitational wave signals.

\begin{figure}[htbp]
    \centering
    \includegraphics[width=8.6cm]{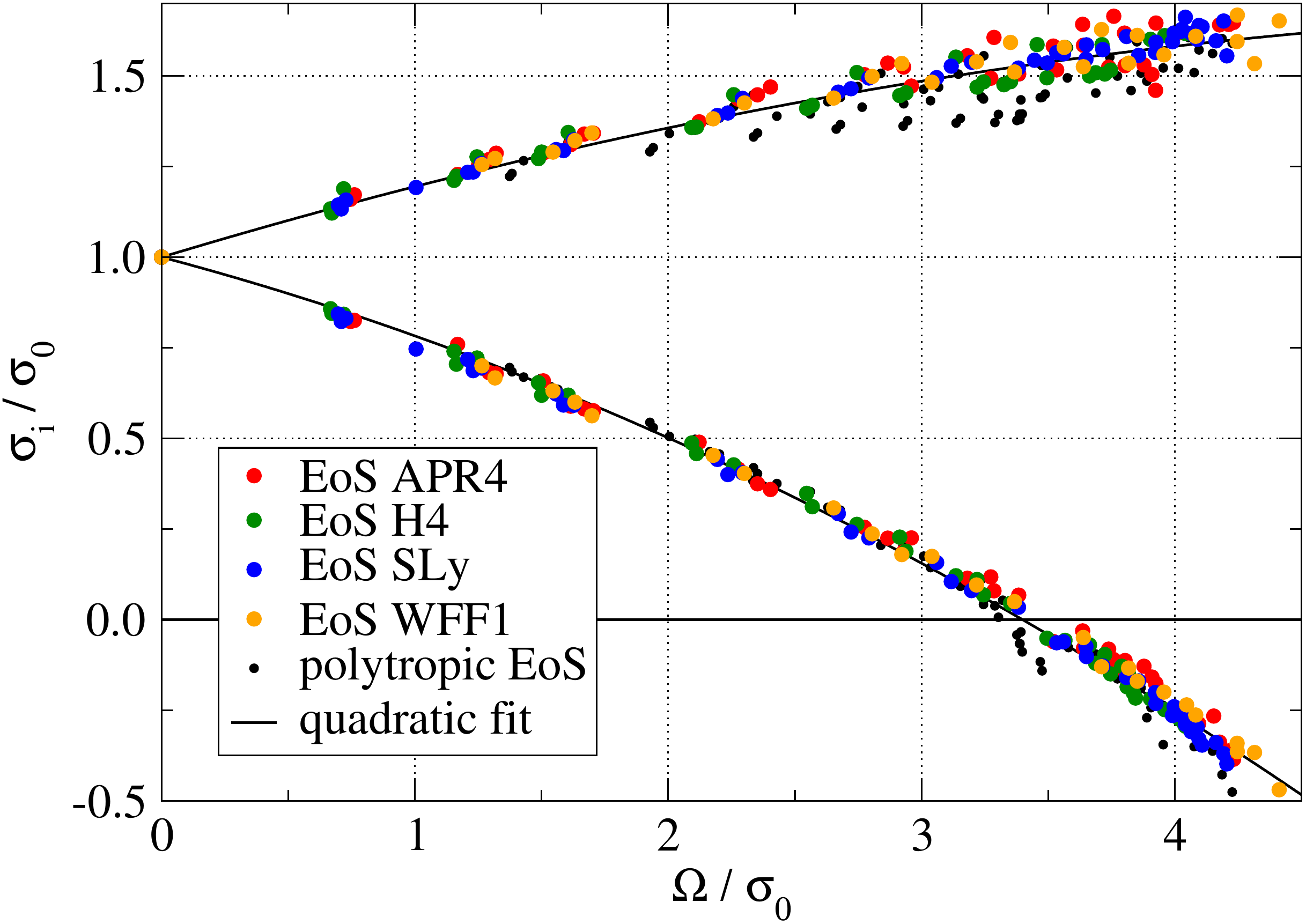}
    \caption{Universal relations for the $l=|m|=2$ $f$-mode frequencies for sequences of constant central energy density as observed in the inertial frame. The graph shows the results from 21 such sequences (three sequences per EoS; three polytropic and four realistic EoSs). The potentially unstable $f$-mode branch displays a strikingly universal behavior; the largest deviations from a quadratic fit occur close to the mass-shedding limit of the sequences, mainly for the polytropic EoSs.}
    \label{fig:sigma-omega-eps}
\end{figure}

More specifically, we will show different universal relations providing accurate estimates for the $f$-mode frequency given some bulk parameters of the star and vice versa. First, we observe a universal behavior of the $f$-mode frequency $\sigma_\text{i}$ as observed in the inertial frame as a function of the star's angular spinning frequency $\Omega$ along sequences of fixed central energy density models when we normalize both frequencies with the $f$-mode frequency $\sigma_0$ of the corresponding non-rotating star. Figure~\ref{fig:sigma-omega-eps} displays this behavior for more than 230 different neutron star models of each the co- and counter-rotating branches of the $f$-mode for seven EoSs and various central energy densities (with corresponding central rest mass densities $\rho_c \in [2.2, 7.3] \rho_0$, where $\rho_0 = 2.7 \times 10^{14}\unit{g/cm}^3$ is the nuclear saturation density); we model the universal behavior using the quadratic function
\begin{align}
    \frac{\sigma_\text{i}}{\sigma_0}
        = 1 + a_1 \left( \frac{\Omega}{\sigma_0} \right)
        + a_2 \left( \frac{\Omega}{\sigma_0} \right)^2 .
    \label{eq:omvsom_model}
\end{align}
The results of a least squares fit are $a_1^\text{u} = -0.193$ and $a_2^\text{u} = -0.0294$ for the potentially unstable branch and $a_1^\text{s} = 0.220$ and $a_2^\text{s} = -0.0170$ for the stable branch of the $f$-mode. The quadratic fit accounts well for the increasing oblateness of the star with its rotation; however, close to the Kepler limit, deviations from this simple model become visible. As this deviation is most pronounced for the less realistic polytropic EoSs, we do not take them into account for the quadratic fits. The root mean square of the residuals is $0.024$ for the counter-rotating branch and $0.048$ for the co-rotating branch.

We point out that our model predicts that the unstable branch of the quadrupolar $f$-mode becomes susceptible to the CFS instability once the angular rotation rate of the star exceeds $\Omega \approx \left(3.4 \pm 0.1\right) \sigma_0$ (when considering sequences of constant central energy density); note that the given uncertainty is a bound, not a confidence interval. This finding regarding the critical value complements the well-known threshold of $T/|W| \approx 0.08 \pm 0.01$ in terms of the ratio of rotational to gravitational potential energy \cite{1998ApJ...492..301S, 1999ApJ...510..854M}, which is confirmed in our simulations and is in contrast to the widely used Newtonian result of $T/|W| \approx 0.14$.

\begin{figure}[htbp]
    \centering
    \includegraphics[width=8.6cm]{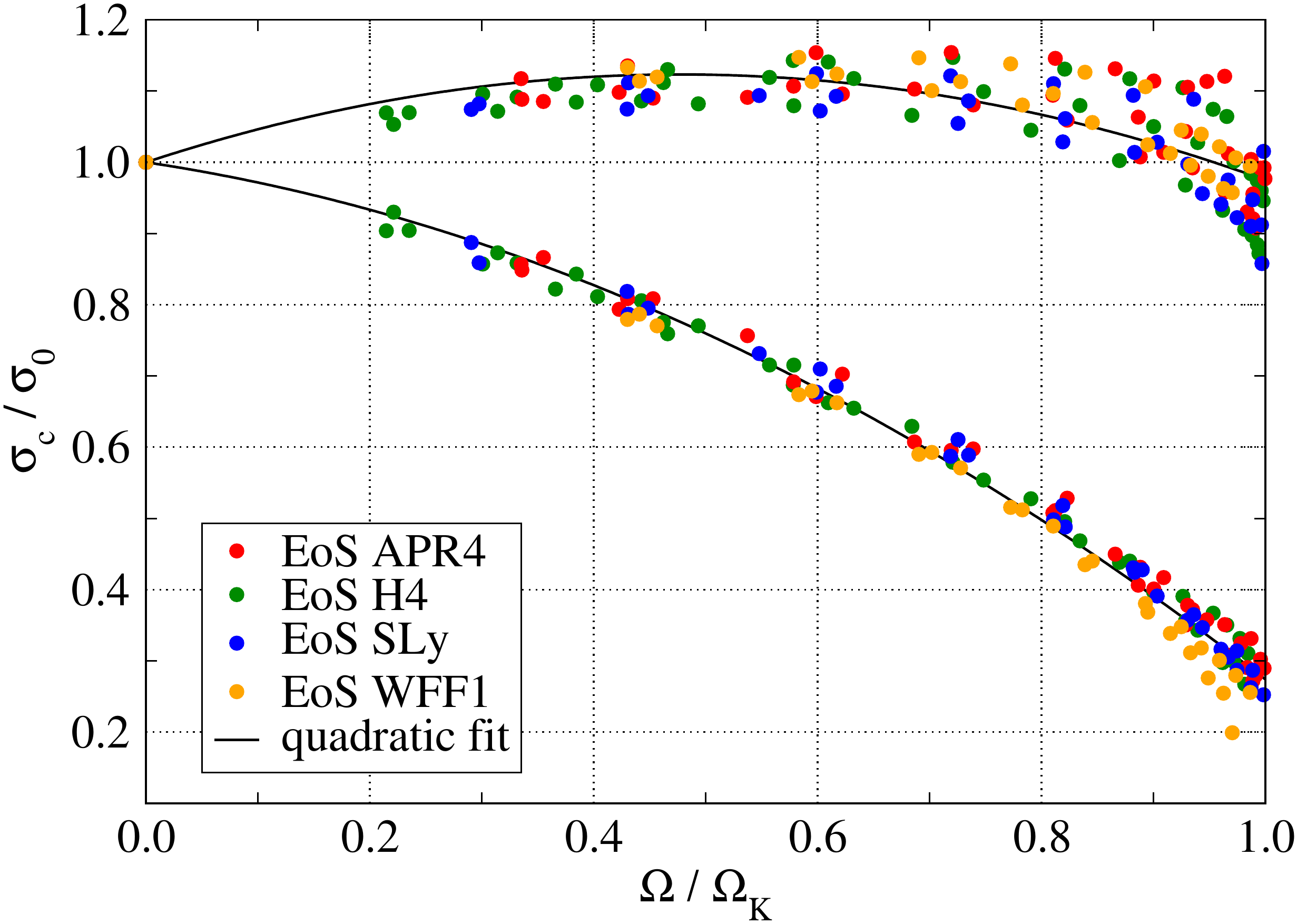}
    \caption{Universal relations for the $l=|m|=2$ $f$-mode frequencies for sequences of constant baryon mass as observed in the comoving frame. The graph shows the results from 12 such sequences (three sequences per EoS). The stable $f$-mode branch displays universal behavior.}
    \label{fig:sigma-omega-Mo}
\end{figure}

The stable branch of the $f$-mode can be fitted more accurately when switching to the comoving frame and considering sequences of constant baryon mass. The frequency $\sigma_\text{c}$ observed in the comoving frame is related to the frequency observed in the inertial frame via $\sigma_\text{c} = \sigma_\text{i} + m\Omega/2\pi$. We show our results for more than 120 different neutron star models using four realistic EoSs in Figure~\ref{fig:sigma-omega-Mo}. We fit our results to the quadratic function
\begin{align}
    \frac{\sigma_\text{c}}{\sigma_0}
        = 1 + b_1 \left( \frac{\Omega}{\Omega_K} \right)
        + b_2 \left( \frac{\Omega}{\Omega_K} \right)^2 ;
    \label{eq:sigma-omega-Mo_model}
\end{align}
note that we use the Kepler velocity $\Omega_K$ to normalize the star's rotation rate in this formula. The results of a least squares fit are $b_1^\text{u} = 0.517$ and $b_2^\text{u} = -0.542$ for the potentially unstable branch (which in the comoving frame exhibits the higher frequencies) and $b_1^\text{s} = -0.235$ and $b_2^\text{s} = -0.491$ for the stable branch of the $f$-mode. The root mean square of the residuals is $0.024$ for the co-rotating branch and $0.051$ for the counter-rotating branch.

In earlier studies for non-rotating models, fitting relations of the form $\sigma_0=\alpha+\beta\sqrt{M_0/R_0^3}$ were derived \cite{1998MNRAS.299.1059A,2011PhRvD..83f4031G}. Here, $\alpha$ and $\beta$ can be estimated for the EoSs that fulfil the constraints at the time of observation while $M_0$ and $R_0$ correspond to the mass and radius of the non-rotating model. Thus, this relation in combination with Eq.~\eqref{eq:omvsom_model} or \eqref{eq:sigma-omega-Mo_model} connects three fundamental parameters of the sequence, \ie mass and radius of the non-rotating member with the spin of the observed model. Obviously, from a single observation of the $f$-mode frequency, one cannot extract these values but can put constraints among the three of them. Any extra observed oscillation frequency, \eg both co- and counter-rotating frequencies or knowledge of some parameters of the star, such as its mass, will place more stringent constraints. 

\begin{figure}[htbp]
    \centering
    \includegraphics[width=8.6cm]{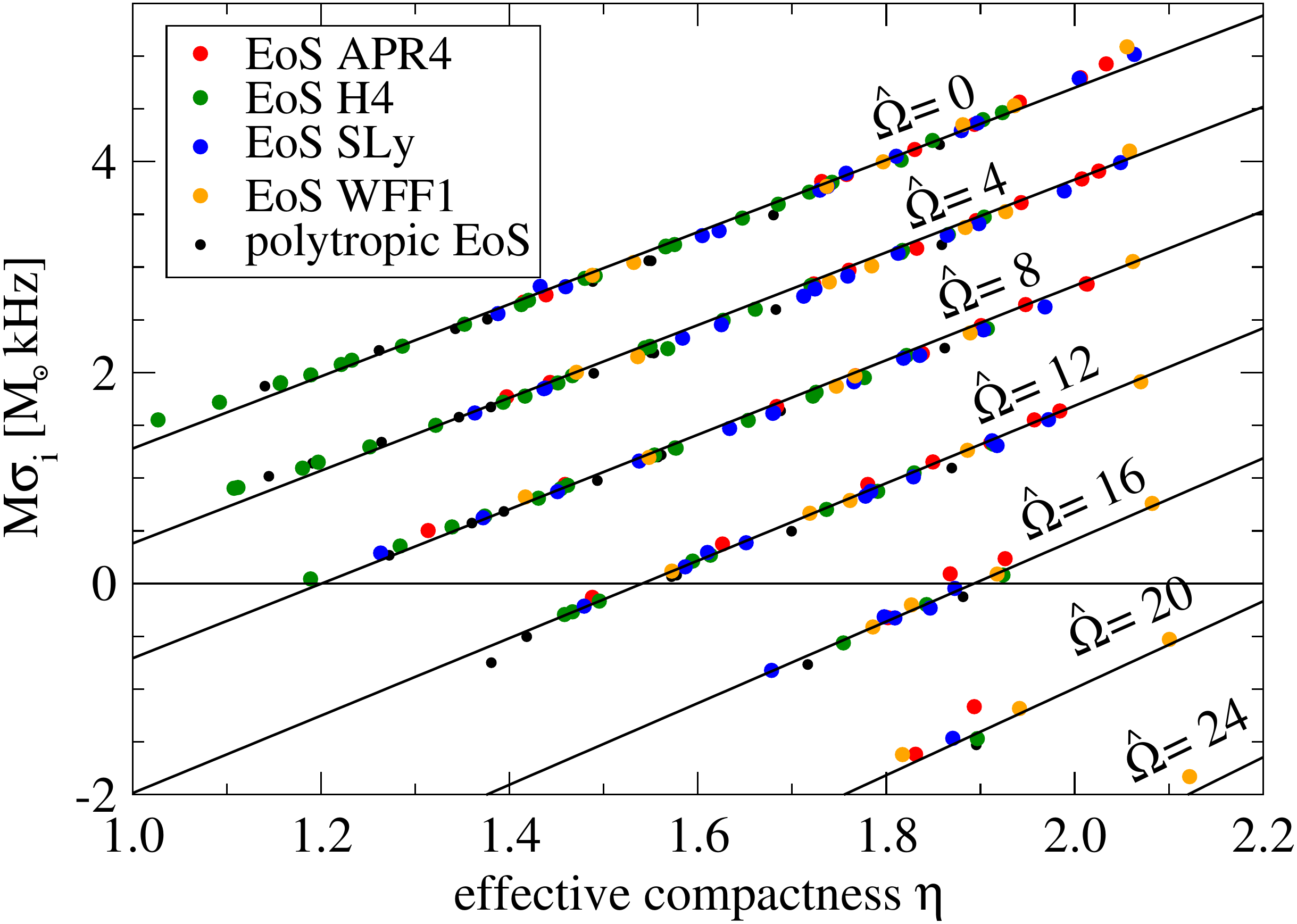}
    \caption{The scaled $f$-mode frequency of the potentially unstable branch in dependence of the effective compactness for different values of $\hat{\Omega} = \bar{M}\Omega/\unit{kHz}$. The straight lines represent the prediction of our fitting formula, cf. Eq.~\eqref{eq:fit-freq-alter}.}
    \label{fig:eta-Msigma}
\end{figure}

Another fitting relation which can easily be implemented in solving the inverse problem is incorporating the \emph{effective compactness} $\eta := \sqrt{\bar{M}^3/I_{45}}$ (which is closely related to the compactness $M/R$), where $\bar{M}:=M/M_\odot$ and $I_{45}:=I/10^{45}\unit{g}\unit{cm}^2$ are the star's scaled gravitational mass and moment of inertia, inspired by \cite{2010ApJ...714.1234L}. We will be guided by the model employed in the Cowling approximation which reproduces the $f$-mode frequency of a particular neutron star from its rotation rate, gravitational mass, and effective compactness \cite{2015PhRvD..92l4004D}. We propose the fitting formula
\begin{align}
    \hat{\sigma_\text{i}} =
    \left(
        c_1 + c_2 {\hat \Omega} + c_3 {\hat \Omega}^2
    \right) +
    \left(
        d_1 + d_3 {\hat \Omega}^2
    \right) \eta,
    \label{eq:fit-freq-alter}
\end{align}
where $\hat{\sigma}_\text{i} := \bar{M}\sigma_\text{i}/\unit{kHz}$ and $\hat{\Omega} := \bar{M}\Omega/\unit{kHz}$; note that we set $d_2 = 0$ as it turns out that this coefficient would be afflicted with a large uncertainty. Using around 100 models based on polytropic as well as around 400 models based on realistic EoSs, the resulting coefficients from a least-squares fit for the counter-rotating branch of the $f$-mode are $(c_1, c_2, c_3)^\text{u}=(-2.14, -0.201, -7.68 \times 10^{-3})$ and $(d_1, d_2, d_3)^\text{u}=(3.42, 0, 1.75 \times 10^{-3})$; for the co-rotating branch, we find the coefficients $(c_1, c_2, c_3)^\text{s}=(-2.14, 0.220, -14.6 \times 10^{-3})$ and $(d_1, d_2, d_3)^\text{s}=(3.42, 0, 6.86 \times 10^{-3})$.
The error in the above reported coefficients is less than 10\,\% and the fitting formula recovers the frequencies with a deviation of less than 20\,\%, with considerably higher accuracy (below 5\,\%) where the $f$-mode frequency is larger than $\approx 500\unit{Hz}$. We show the obtained frequencies along with the predictions from our proposed fitting formula for a few select values of $\hat{\Omega}$, spanning the parameter space up to the Kepler limit, in Fig.~\ref{fig:eta-Msigma}.
 
Qualitatively, our coefficients for the counter-rotating branch agree in order of magnitude with those in \cite{2015PhRvD..92l4004D} in the Cowling approximation; comparing the special case of no rotation, $\hat{\Omega} = 0$, our fitting formula yields roughly $20\,\%$ lower frequencies in our fully general relativistic setup, which is in accordance with expectations.

The lines of constant $\hat{\Omega}$ in Figure~\ref{fig:eta-Msigma} may give the impression that the CFS instability operates more easily in stars with low (effective) compactness, seemingly in contrast to the finding that post-Newtonian effects tend to enhance this instability \cite{1992ApJ...385..630C}. This paradox can be resolved by noting that relativistic effects mainly shift the $f$-mode frequency to lower values while the inclination of the lines of constant $\hat{\Omega}$ is largely unaltered (cf. Figure~3 in \cite{2015PhRvD..92l4004D}). Furthermore, while stars of lower (effective) compactness may reach the neutral point of the $f$-mode indeed at a lower rotation rate, this happens considerably closer to the Kepler limit (if at all) than it would do in more compact stars.

The fitting formula \eqref{eq:fit-freq-alter} has the advantage that it does not rely on specifically defined sequences of neutron stars, along which a particular property is held constant. For example, Eq.~\eqref{eq:sigma-omega-Mo_model} depends on the $f$-mode frequency $\sigma_0$ of the (in a very particular fashion) corresponding non-rotating configuration, which may not even exist in some cases (\eg for supramassive neutron stars supported by rotation); the latter model, cf.~Eq.~\eqref{eq:fit-freq-alter}, is satisfied with bulk properties of the star of which we want to know the oscillation frequency and vice versa. Another benefit of this formulation is that (as demonstrated in \cite{2015PhRvD..92l4004D}) a similar formula can be derived for higher multipoles, \ie $l \ge 3$. Fitting formula \eqref{eq:fit-freq-alter} can be useful in imposing further constraints on the parameters of the postmerger objects since it combines the mass and spin of the resulting object with the $f$-mode frequency and, via $\eta$, the moment of inertia $I$ or the compactness $M/R$. Thus, the latter two can be further constrained by an observation of an $f$-mode signal, as mass and potentially spin can be extracted from the premerger and early postmerger analysis of the signal. The situation becomes more attractive if both co- and counter-rotating modes or other combination of modes are observed since only the mass of the postmerger object will be needed to constrain its parameters by using only the asteroseismological relations \cite{2011PhRvD..83f4031G,2015PhRvD..92l4004D,2020PhRvD.101h4039}. This will be an independent yet complementary constraint in the estimation of the radius in addition to those based on the Love numbers \cite{2018PhRvD..98h4061D,2018PhRvL.121p1101A,2018PhRvL.121i1102D,2018ApJ...852L..29R}.

\begin{figure}[htbp]
    \centering
    \includegraphics[width=8.6cm]{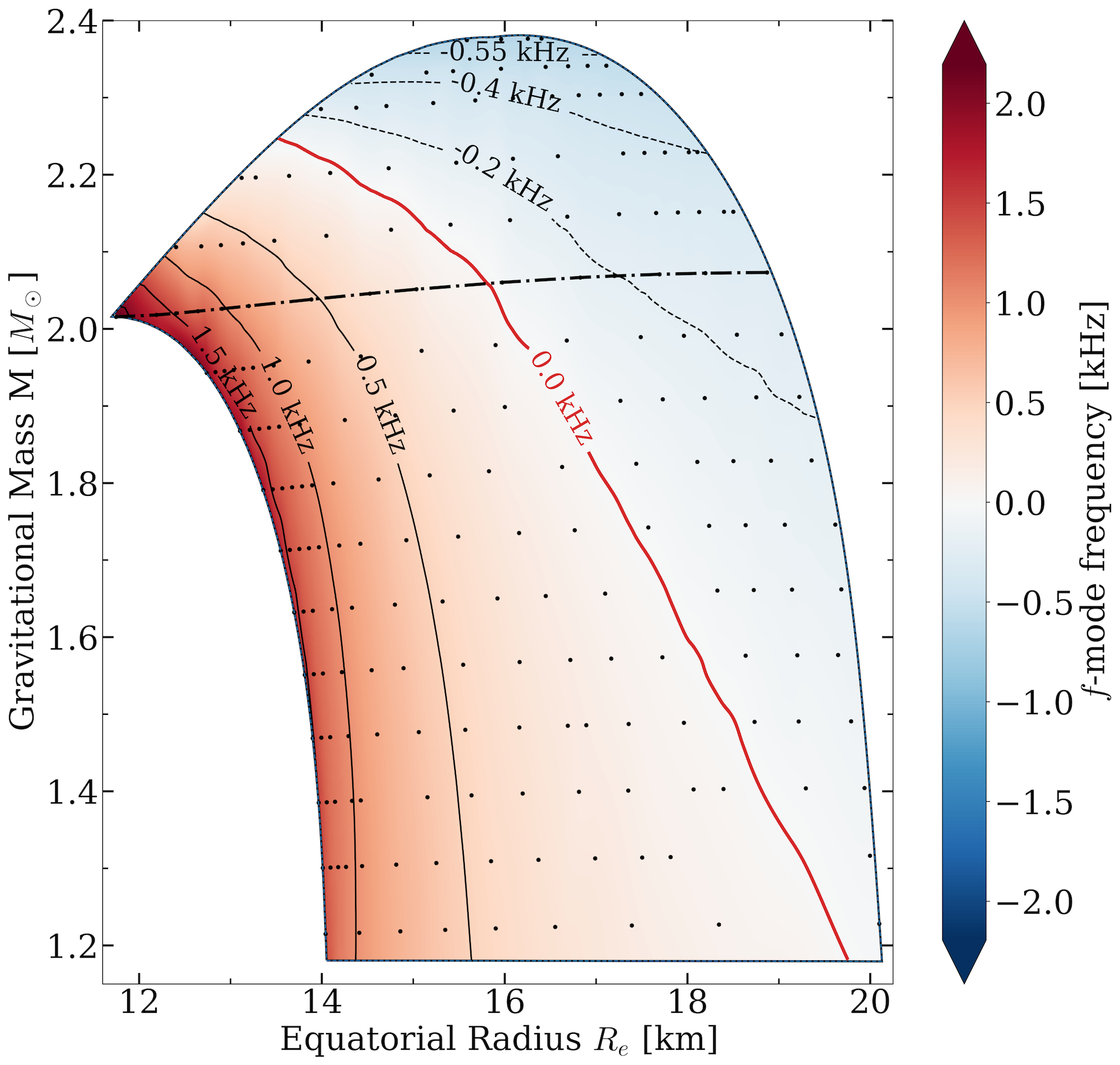}
    \caption{The frequency of the ${}^2f_2$-mode for the entire EoS H4 is displayed color coded and some contour lines are shown. Each black dot indicates a neutron star model for which we have calculated its nonaxisymmetric mode frequencies. All neutron stars located above the (nearly horizontal) dash-dotted line are supramassive. The red (thick) contour line at $0.0\unit{kHz}$ separates the stable models from those that are susceptible to the CFS-instability.}
    \label{fig:H4fmode}
\end{figure}

Finally, as a graphical illustration of the behavior of the $f$-mode frequency across the entire parameter space of stable equilibrium models, we present in Figure~\ref{fig:H4fmode} the frequency of the counter-rotating branch as obtained from the time evolutions exemplary for EoS H4 (expecting qualitatively similar results for other EoSs). We constructed 214 neutron star models across the entire $M$-$R_e$ plane which has four distinct boundaries: the static limit, the mass-shedding limit, and the limit of stability with respect to quasi-radial perturbations; furthermore, in line with current theory and observations \cite{2018MNRAS.481.3305S,2015ApJ...812..143M}, we limit ourselves to neutron stars with masses $M \gtrsim 1.17 M_\odot$.

\emph{Conclusions and outlook.}---We report the first extraction of frequencies
of the $l=|m|=2$ $f$-mode of general relativistic, rapidly rotating neutron stars \emph{without} the commonly used \emph{slow-rotation} or \emph{Cowling approximation} to an extent that allows us to generalize the findings into universal relations. This concludes a long-standing open problem, building upon the effort from numerous studies throughout the past five decades. 

We provide different universal relations for the frequencies of the $l=|m|=2$ $f$-modes of uniformly rotating neutron stars at zero temperature which are independent of the EoS; the proposed formulae are calibrated to several hundred neutron star models that are constructed using both polytropic and realistic EoSs and are scattered across the entire parameter space of equilibrium solutions. Such universal relations will be an essential piece in the asteroseismological toolkit once the third generation GW observatories will be able to pick up the ring-down and fluid ringing signal following the merger of a binary neutron star system; they allow to solve the inverse problem, leading to significantly tighter constraints for mass and radius of the postmerger object. For this task, it is elementary to have a smorgasbord of universal relations at hand, which allows to make a practical choice, depending on which observables are available or in which of the star's properties one is interested. We will extend the present list of such universal relations in future articles, utilizing different combinations of bulk properties of the star; while we may obviously be (and already have been) inspired by previously published fitting formulae that were derived using different approximative frameworks, we need to be open-minded about models involving novel combinations of observables.

We also report the discovery of an accurate estimate for the onset of the CFS-instability when the $f$-mode frequency of the non-rotating member of the family is known and verified the corresponding critical value of $T/|W|$.

A natural extension of the present Letter will be a more comprehensive investigation of the spectrum of neutron stars (i.e. higher multipole $f$-modes, low $p$-modes and $g$-modes as well as $w$-modes) which may be excited in different astrophysical processes. Furthermore, we are going to extend our code to account for differentially rotating neutron stars and hot EoSs that are particularly relevant for nascent neutron stars or postmerger configurations in the immediate aftermath of a binary merger, both of which will have a considerable impact on the vibration frequencies or the onset of the CFS-instability (and via two further scaling parameters also on the universal relations) during a very short but dynamic interval of their lives.

This work was supported by DFG research Grant No. 413873357. A part of the computations were performed on Trillian, a Cray XE6m-200 supercomputer at UNH supported by the NSF MRI program under Grant No. PHY-1229408.

\end{document}